# Coupling model of metallic target ablation-plasma evolution-radiation under nanosecond laser irradiation*

ZHOU Ying[1], WU Jian[1,*], SUN Hao[1], LI Jinghui[1], LI Xiaoxuan[1], HUANG Shuzhi[1], HE Jiayao[1], LIU Xingyu[1], HANG Yuhua[2], PEI Cuixiang[3], LI Xingwen[1]

1. State Key Laboratory of Electrical Insulation and Power Equipment, Xi’an Jiaotong University, Xi’an 710049, China
2. Suzhou Nuclear Power Research Institute Co., Ltd., Suzhou 215004, China
3. School of Aerospace Engineering, Xi’an Jiaotong University, Xi’an 710049, China

**Abstract**

The interaction of nanosecond laser pulses with metallic materials involves multiple complex physical processes, and constructing a self-consistent model capable of uniformly describing all stages remains a significant challenge. This work establishes a multi-physics coupling model for pure iron, encompassing laser energy deposition, solid-liquid phase transition, gas-liquid interfacial kinetic transport, plasma expansion and ionization, and spectral radiation. The numerical solution adopts a partition method, utilizing an implicit compact difference scheme for the target area and a Mac-Cormack explicit scheme for the ambient atmosphere, to simulate the ablation dynamics. The simulations elucidate the emergence of plasma shielding and its inhibitory effect on the evaporation process, thereby confirming that 81.6% of the early-stage ablation products are transported through a supersonic expansion mode. The model successfully captures the complete evolution of the plasma plume from a high-temperature, highly ionized state (dominated by $Fe^{3+}$) to a low-temperature, neutral atomic state (dominated by $Fe^{0}$). Based on this, spectral calculations demonstrate the dynamic evolution of radiative characteristics from an early stage featuring a “strong continuum background dominated by ion lines” to a later stage

---

† Corresponding author E-mail: jxjawj@mail.xjtu.edu.cn

The first author E-mail: yingzhou@stu.xjtu.edu.cn

where “the continuum attenuates, atomic lines become prominent, and self-absorption appears”. The emergence of self-absorption proves the ability of the model to effectively capture the optical thickness effects arising from spatial inhomogeneity within the plasma. Through systematic comparison between experimentally measured spectra and calculated results from the PrismSPECT and NIST LIBS spectral programs, the model presented here achieves the highest comprehensive scores in quantitative evaluations of multiple channels. This validates the necessity and superiority of the full-chain self-consistent modeling approach over traditional methods relying on spatial averaging or optically thin approximation, especially in describing plasma inhomogeneity and radiation transport. It also provides a numerical simulation framework for applications such as laser processing parameter optimization, quantitative spectroscopic analysis, and the design of novel plasma light sources.

# 1 Introduction

The interaction between nanosecond lasers and metallic materials serves as a core physical process in numerous fields, including high-energy-density physics, high-precision manufacturing, plasma light sources, and material modification. Consequently, it has remained a focal point of research both domestically and internationally. Since Maiman constructed the first ruby laser at Bell Labs in the United States in 1960, laser technology has been widely applied in surface etching, cleaning, thin-film deposition, laser-induced breakdown spectroscopy (LIBS), and optical diagnostics[1]. However, this multiphysics system exhibits highly non-equilibrium and strongly coupled characteristics[2]. It involves multi-scale, cross-stage physical couplings, such as laser energy absorption, material heat conduction, phase-change evaporation, plasma formation and expansion, particle ionization and recombination, and radiative transport[3,4]. These processes span time scales from nanoseconds to microseconds and spatial scales from nanometers to centimeters. Their nonlinear features and complex coupling among multiple parameters pose significant challenges for both experimental diagnostics and theoretical modeling[5,6]. Therefore, constructing a numerical model that uniformly describes the entire chain of laser-solid-vapor-plasma-radiation is a crucial theoretical support and a major research challenge. Such a model is essential for understanding laser-matter interactions, guiding the optimization of experimental parameters, and achieving controllable processing.

Over the past few decades, numerous studies domestically and internationally have made significant progress in modeling single or partial stages of this process[7-11]. For instance, the classical laser ablation theory proposed by Anisimov[12] effectively describes the energy absorption and evaporation processes of solid materials under intense laser irradiation. The Knudsen layer model established by Knight[13] reveals the fundamental laws of non-equilibrium transport at the gas-liquid interface. With the development of computational physics, various general-purpose programs for simulating laser-low-temperature plasma interactions have subsequently been developed.

In terms of fluid modeling, the FLASH code developed by Fryxell et al.[14] is currently one of the most widely used open-source hydrodynamic frameworks. It can simultaneously handle multiphysics processes such as laser energy deposition, shock wave propagation, and spatial confinement. Guthikonda et al.[15] used FLASH to simulate the evolution of laser-induced plasma and its shock waves under spatial confinement. Their results indicated that the primary shock wave doubles the temperature and density of the plasma through multiple reflections and compression, which helps enhance LIBS signal intensity. The flux-corrected two-dimensional hydrodynamic code POLLUX, developed by Pert et al.[16,17], combines the CHART-D equation of state with the Thomas-Fermi electron model. This approach enables the extension of the equation of state for material ablation phase transitions at low laser fluences ($\leq$10 J/cm$^2$) and simulates the spatiotemporal evolution of early plasma expansion. Hill and Wagenaars[18] used POLLUX to systematically investigate the influence of material parameters of six metals on the temperature and particle number density of laser-ablated plasma.

In the realm of radiation-hydrodynamics modeling, the RALEF-2D code developed by Basko et al.[19] employs an arbitrary Lagrangian-Eulerian (ALE) scheme. It can self-consistently describe the dynamic evolution and radiative characteristics of laser-driven plasma. Torretti et al.[20] used RALEF-2D to simulate plasma generated by laser irradiation of tin droplets. They demonstrated that extreme ultraviolet radiation near 13.5 nm primarily originates from multi-excited state transitions of $Sn^{11+}$-$Sn^{14+}$ rather than single-electron transitions. The Prism series of high-energy-density physics modules developed by MacFarlane et al.[21-23] enables coupled solutions of multi-group radiative transport and hydrodynamics. It also facilitates spectral calculations under local thermodynamic equilibrium (LTE) or collisional-radiative (CR) models. Joshi et al.[24] inferred the electron temperature and electron number density of laser-ablated silicon plasma based on synthetic spectra calculated by PrismSPECT. They analyzed the influence of pulse width on plasma ionization under constant laser fluence. The NIST LIBS code[25], developed by the National Institute of Standards and Technology (NIST) in the United States, can calculate line spectra of low-temperature plasma with mixed components under LTE. Veis et al.[26] simulated spectra using the NIST LIBS code and demonstrated that introducing spectral lines in the vacuum ultraviolet region (<200 nm) improves the accuracy of electron temperatures calculated by the Saha-Boltzmann method. This enhancement thereby improves the detection capability of LIBS for light elements such as Si, B, C, and S.

Although existing models and codes have achieved significant progress in modeling laser ablation and plasma dynamics, most studies still employ independent physical assumptions or empirical parameters for different stages. For example, in the early stage where solid, liquid, gas, and plasma phases coexist, the latent heat of phase change, transport at the evaporation front interface, and subsequent plume ionization are often partially ignored or treated in isolation[27,28]. While hydrodynamic and plasma dynamic modules can describe the radiative transport and dynamic evolution of plasma over large spatiotemporal scales, they often neglect laser energy deposition within the solid. Instead, they typically apply simplified absorption layer boundary conditions in the solid region[29,30]. Furthermore, these models largely rely on high-temperature equations of state. Although such equations effectively handle thermodynamic properties in rarefied, high-temperature regions, their applicability in the solid/phase-transition region is significantly limited. Consequently, they struggle to accurately reflect microscopic processes such as local energy density changes and latent heat release[31-33]. Additionally, coupling between different modules is often implemented via unidirectional data transfer in practice. This non-self-consistent solution method makes it difficult to maintain energy conservation and physical continuity.

In recent years, the academic community has gradually initiated research on cross-stage full-chain modeling. For example, the ALE3D code developed by Haxhimali et al.[34,35] employs a hybrid finite element/finite volume method with unstructured meshes. It integrates multiple physical modules, including solid mechanics, heat transfer, chemical reactions, component diffusion, and compressible/incompressible flow. However, it does not yet include functions for radiative transport and spectral calculations. The multi-group radiation hydrodynamics code RHDLPP, developed by Min et al.[36,37], achieves coupling between solid and plasma regions by introducing heat transfer and Knudsen layer boundaries. They also developed a spectral post-processing module, RHDLPP-SpeIma3D[38], for radiative transport simulation and transient/spatiotemporal integrated spectral output. However, this model describes the flow field using Euler equations that ignore diffusion, viscosity, and heat transfer. This limitation restricts its ability to characterize viscous dissipation and energy transport processes. Research by Gornushkin et al.[9] indicates that transport mechanisms such as viscosity, diffusion, and heat conduction have significant and spatiotemporally non-uniform effects on plasma density distribution. Ignoring these physical processes may affect the accurate assessment of atmospheric effects and plasma evolution behavior.

Based on the current research status and existing limitations, this study uses pure iron as the research object. We attempt to construct a multiphysics coupled computational model covering the entire process. This includes laser energy deposition, phase-change evaporation, gas-liquid interface transport, plasma expansion and ionization, component diffusion, thermal/viscous transport, and spectral radiation. To verify the reliability of the computational results, we quantitatively compare the calculated spectra with experimental measurements and results from the PrismSPECT and NIST LIBS codes. This work aims to provide a complete and reliable numerical simulation platform for understanding the physical processes of laser-matter interaction.

# 2 Computational Model

The interaction between a nanosecond laser and a solid target in a background atmosphere is a complex transient process involving multiphysics coupling. Using an iron target as an example, this study establishes a complete computational model and numerical solution framework ranging from laser ablation deposition to spectral radiation. The model sequentially describes laser ablation of the target material, kinetic transport at the gas-liquid interface, plasma expansion and ionization, component diffusion and thermal/viscous transport, and final spectral radiation.

## 2.1 Nanosecond Laser Ablation Model Considering Solid-Liquid Phase Change

During nanosecond laser ablation, temperature changes induced by laser incidence and their effects on material phase change and ablation are critical. To accurately simulate this process, the laser ablation model employed in this study must comprehensively consider solid-liquid phase change, heat conduction, and evaporation. It is important to note that this model primarily targets single-pulse or low-repetition-rate conditions. These conditions typically imply that the pulse interval is much longer than the material's thermal relaxation time. For example, the thermal relaxation time of pure iron with a spot size of hundreds of micrometers is approximately 440 μs. Under such conditions, inter-pulse thermal accumulation effects can be ignored, and the action of each pulse can be treated as independent.

First, to avoid numerical instability, the temperature field in the solid-liquid phase change region requires a smooth transition. This study employs a Gaussian distribution operator for this purpose[39]:

$$\delta(T-T_m)=\frac{1}{\sqrt{2\pi}\Delta}\mathrm{e}^{\frac{-(T-T_m)^2}{2\Delta^2}} \tag{1}$$

Here, $\delta(T - T_{\mathrm{m}})$ represents the temporal variation of temperature within the solid-liquid phase change region. $T_{\mathrm{m}}$ denotes the melting point of the material. The parameter $\Delta$ controls the width of the phase change zone and determines the upper ($T_{\mathrm{mH}}$) and lower ($T_{\mathrm{mL}}$) temperature bounds.

Consequently, this operator divides the heat transfer process into three intervals based on the temperature distribution domain: the solid phase ($T<T_{\mathrm{mL}}$), the liquid phase ($T>T_{\mathrm{mH}}$), and the solid-liquid coexistence region ($T_{\mathrm{mL}}<T<T_{\mathrm{mH}}$). By incorporating the physical property parameters specific to each region, the governing heat transfer equation is expanded as follows

$$\begin{cases} \rho_i c_i(\frac{\partial T}{\partial t}-u\frac{\partial T}{\partial x})=\lambda_i\frac{\partial^2 T}{\partial x^2}+(1-R)\alpha I(t)e^{-\alpha x}, & T_{mH}<T(i=s);T<T_{mL}(i=l) \\ \rho_s\left[c_s+\Delta H_m\delta(T-T_m)\right](\frac{\partial T}{\partial t}-u\frac{\partial T}{\partial x})=\lambda_s\frac{\partial^2 T}{\partial x^2}+(1-R)\alpha I(t)e^{-\alpha x}, & T_{mL}<T<T_{mH} \end{cases} \tag{2}$$

Here, $\rho_i$, $c_i$, and $\lambda_i$ denote the mass density, specific heat capacity, and thermal conductivity for the solid and liquid phases, respectively; $x$ represents the spatial coordinate; $\alpha$ is the optical absorption coefficient; $I(t)$ denotes the time-dependent incident laser irradiance; $R$ is the reflectivity of the material surface to incident laser; $\Delta H_{\mathrm{m}}$ represents the latent heat of fusion; and $u$ indicates the evaporation rate at the material surface. The evaporation rate $u$ is jointly determined by the saturated vapor pressure $P_{\mathrm{s}}$ and the surface temperature $T_{\mathrm{s}}$ using the Clausius-Clapeyron relation and the Hertz-Knudsen model.

At the material surface, the incident laser energy affects the surface temperature and evaporation rate through heat flux transfer. At $x = 0$, the heat flux boundary condition coupled with the evaporation effect is described as

$$u\left(\rho\Delta H_e+\rho U+\rho v^2/2+P\right)-\lambda_s\frac{\partial T}{\partial x}=(1-R)I(t)\mathrm{e}^{-\alpha x}, \quad x=0 \tag{3}$$

In this equation, the terms on the left-hand side independently describe the changes in latent heat during evaporation($u\rho\Delta H_e$), internal energy stored by vapor expansion ($\rho U$), macroscopic kinetic energy ($\rho v^2/2$), work done by expansion pressure ($P$), and conductive heat transfer ($\lambda_i\partial T/\partial x$). The right-hand side represents the deposition of laser irradiance on the material surface.

Subsequently, as the material surface temperature continues to rise, the evaporation effect gradually becomes dominant. This process influences the velocity distribution and mass transport at the gas-liquid interface.

## 2.2 Velocity Distribution and Mass Transport at the Gas-Liquid Interface

To simulate the dynamic behavior of the gas-liquid interface, this study introduces the Knudsen layer as a kinetic discontinuity at the interface. Within this region, the velocity distribution of vapor particles initially follows a half-Maxwellian distribution. As inter-particle collisions increase, the distribution gradually approaches an equilibrium Maxwellian distribution. Finally, it restores to an equilibrium Maxwellian distribution outside the Knudsen layer.

Under this definition, the particle velocity distribution function $f(x, \boldsymbol{v})$ within the layer can be expressed as a weighted sum of the velocity distribution functions on both sides[40]:

$$f(x,\mathbf{v}) = \alpha(x) f_1(\mathbf{v}) + [1-\alpha(x)] f_2(\mathbf{v}) \tag{4}$$

Here, $\alpha(x)$ denotes the spatial transition weight function, satisfying $\alpha(0) = 1$ and $\alpha(\infty) = 0$. $f_1(\boldsymbol{v})$ represents the velocity distribution function of particles newly released near the liquid surface, while $f_2(\boldsymbol{v})$ denotes the Maxwellian velocity distribution in equilibrium on the vapor side:

$$f_1(\mathbf{v}) = \begin{cases} n_s \left( \dfrac{m}{2\pi k T_s} \right)^{3/2} \exp\left( -\dfrac{m v_s^2}{2 k_B T_s} \right), & v_k > 0 \\ \beta f_2(\mathbf{v}), \quad v_k < 0 \end{cases} \tag{5}$$

Here, the positive branch of $f_1(\boldsymbol{v})$ follows a half-Maxwellian distribution for the liquid surface temperature $T_s$. The negative branch ensures momentum conservation across the interface by adjusting the coefficient $\beta$ for the reflux particle flux. $k_B$ denotes the Boltzmann constant; $T_k$ represents the vapor temperature; $n_s$ and $n_k$ are the particle number densities near the liquid surface and in the vapor, respectively; and $v_k$ is the mean vapor velocity.

Substituting the aforementioned velocity distribution into the mass, momentum, and energy conservation relations for both sides of the layer yields the temperature ratio, density ratio, and reflux coefficient across the gas-liquid interface:

$$
\begin{cases}
\dfrac{T_k}{T_s} = \left( \sqrt{1 + \pi\left(\dfrac{\gamma_k - 1}{\gamma_k + 1}\dfrac{m}{2}\right)^2} - \sqrt{\pi}\,\dfrac{\gamma_k - 1}{\gamma_k + 1}\dfrac{m}{2} \right)^2 \\
\dfrac{\rho_k}{\rho_s} = \sqrt{\dfrac{T_s}{T_k}}\left[ \left( m^2 + \dfrac{1}{2} \right) e^{m^2} \mathrm{erfc}(m) - \dfrac{m}{\sqrt{\pi}} \right] + \dfrac{1}{2}\dfrac{T_s}{T_k}\left[ 1 - \sqrt{\pi} m e^{m^2} \mathrm{erfc}(m) \right] \\
\beta = \left[ \left( 2m^2 + 1 \right) - m\sqrt{\dfrac{\pi T_s}{T_k}} \right] e^{m^2} \dfrac{\rho_s}{\rho_k}\sqrt{\dfrac{T_s}{T_k}}
\end{cases}
\tag{6}
$$

Here, $\rho_s$ and $\rho_k$ denote the liquid surface and vapor densities, respectively, while $\gamma_k$ represents the specific heat ratio of the vapor. The dimensionless Mach number is defined as $m = u_k \big/ \sqrt{2 R_s T_k}$, where $R_s$ is the gas constant. The complementary error function is given by $\mathrm{erfc}(m) = 2\big/\sqrt{\pi}\int_m^{\infty} \exp\left(-v^2\right) \mathrm{d}v$.

Eq. (6) indicates that the physical quantities on both sides of the interface are determined by the free variable, namely, the dimensionless Mach number $m$ at the outer exit of the layer. When the vapor outside the layer escapes at subsonic speed, the Rankine-Hugoniot conservation relations describing ideal gas shock waves are introduced to provide the determinate conditions:

$$
\frac{P_2}{P_1} = 1 + \gamma_1 M_2 \frac{a_2}{a_1}\left[ \frac{1+\gamma_1}{4} M_2 \frac{a_2}{a_1} + \sqrt{1 + \left( \frac{1+\gamma_1}{4} M_2 \frac{a_2}{a_1} \right)^2} \right]
\tag{7}
$$

Here, $P_1$ and $P_2$ denote the pressures upstream (static gas side) and downstream (vapor side) of the shock wave, respectively; $\gamma_1$ represents the specific heat ratio of the gas medium; $M_2$ is the Mach number at the Knudsen layer exit; and $a_1$ and $a_2$ correspond to the speeds of sound in the gas and vapor phases, respectively.

Conversely, when vapor outside the layer flows back to the liquid surface, forming a condensation counterflow, the mass fluxes across the Knudsen layer are supplemented by an additional kinetic equilibrium condition:

$$
n_k u_k = \frac{P_s - P_k}{\left( 2\pi m_v k_B T_s \right)^{1/2}}
\tag{8}
$$

Here, $n_k$ denotes the number density of vapor outside the layer, $P_s$ and $P_k$ represent the pressures at the liquid surface and in the vapor phase, respectively, and $m_v$ is the mass of a single vapor molecule.

Solving these equations simultaneously yields the initial distributions of pressure,

temperature, velocity, and mass density on the vapor side. This approach facilitates the transitional solution from the ablation region to the flow field distribution in the vapor region.

## 2.3 Plasma Expansion and Ionization Distribution

During the calculation of the flow field on the vapor side, the generated vapor rapidly absorbs laser energy through inverse bremsstrahlung and photoionization effects. This process forms a plasma consisting of electrons, ions, and neutral particles. To describe this phenomenon, we introduce a set of viscous flow equations that account for laser absorption, species diffusion, self-generated pressure gradient driving forces, radiation losses, and thermal losses:

$$\begin{cases} \dfrac{\partial \rho}{\partial t} = -\dfrac{\partial \rho v}{\partial x} \\ \dfrac{\partial \rho_i}{\partial t} = -\dfrac{\partial \rho_i v}{\partial x} + \dfrac{\partial}{\partial x}(\rho D_i \dfrac{\partial w_i}{\partial x}), \quad i = 1,2,...,k \\ \dfrac{\partial (\rho v)}{\partial t} = -\dfrac{\partial}{\partial x}\left(\rho v^2 + P - \tau_{xx}\right) \\ \dfrac{\partial}{\partial t}\left(\rho\left(U + \dfrac{1}{2}v^2\right)\right) = -\dfrac{\partial}{\partial x}\left(\rho\left(U + \dfrac{1}{2}v^2\right)v + Pv + q - \tau_{xx}v\right) + (\alpha_{\mathrm{IB}} + \alpha_{\mathrm{PI}})I - \varepsilon_{\mathrm{rad}} \end{cases} \tag{9}$$

Here, $\rho$ denotes the total density; $\rho_i$, $D_i$, and $w_i$ represent the density, diffusion coefficient, and mass fraction of the $i$-th component, respectively; $v$ is the flow velocity; $P$ is the pressure; $\tau_{xx}$ is the shear stress; $U$ is the internal energy per unit mass; $q$ is the heat flux; $\alpha_{\mathrm{IB}}$ and $\alpha_{\mathrm{PI}}$ denote the inverse bremsstrahlung absorption coefficient and the photoionization absorption coefficient, respectively, with calculation methods detailed in Ref.[41]; $\varepsilon_{\mathrm{rad}}$ represents the radiative energy loss, which is calculated here using bremsstrahlung losses as a representative mechanism. Furthermore, the ideal gas equation of state links particle mass, temperature, and pressure.

In modeling the ionization process, this study adopts a framework based on local thermodynamic equilibrium. This assumption holds during the main phase of plasma evolution and in the near-field region. It applies to nanosecond pulse ablation scenarios with peak irradiance on the order of $10^8$~$10^{10}$ W/cm$^2$. The number density of particles in each ionization state is calculated using the Saha equation. This approach, combined with charge and mass conservation, closes the system of equations:

$$\frac{x_{i+1} n_e}{x_i} = 2\left[\frac{m_e c^2 k_B T}{2\pi (\hbar c)^2}\right]^{\frac{3}{2}} \frac{Z_{i+1}(T)}{Z_i(T)} e^{-\frac{E_i}{k_B T}} = F_{i+1}(T) \qquad i = 0, 1, \ldots, (Z-1) \tag{10}$$

Here, $x_i$ denotes the mass fraction of ions at the $i$-th level, $n_e$ represents the electron number density, $m_e$ is the electron mass, $c$ is the speed of light, $\hbar$ is the reduced Planck constant, and $E_i$ is the ionization energy at the $i$-th level.

## 2.4 Calculation Models for Diffusion Coefficients and Thermal/Viscous Transport Coefficients in Air Atmosphere

Given that ambient air at standard temperature and pressure frequently serves as the background gas in practical laser processing, measurement, and diagnostic applications, we simplify air to a mixture of 78% $N_2$ and 22% $O_2$ to facilitate computation. This section subsequently introduces the calculation models for the diffusion and transport coefficients in an air atmosphere.

The diffusion coefficient governs substance dispersion within the flow field, thereby influencing the mixing uniformity of metal vapors with the background gas and the rate of energy transfer. When three or more components exist in the atmosphere, Wilke introduced an effective diffusion coefficient based on Maxwell-Stefan theory to describe the migration rate of a specific component relative to the overall mixture. The resulting multicomponent diffusion coefficient is given as[42]

$$D_v = \frac{1-x_v}{\sum\limits_{v\neq p}\frac{x_p}{D_{vp}}}, D_{vp} = \frac{2}{3}\left(\frac{k_B T}{\pi}\right)^{\frac{3}{2}}\left[\frac{1}{2}\left(\frac{1}{m_v}+\frac{1}{m_p}\right)\right]^{\frac{1}{2}}\frac{4}{P\left(d_v+d_p\right)^2} \tag{11}$$

Here, $x_v$ and $x_p$ denote the mole fractions of components $v$ and $p$, respectively; $m_v$ and $m_p$ represent the molecular masses; $d_v$ and $d_p$ indicate the effective molecular radii; $P$ is the total system pressure; and $D_{vp}$ is the binary diffusion coefficient of component $v$ relative to each other component, which can be determined via Fick's law.

Furthermore, transport coefficients (viscosity and thermal conductivity) significantly influence the flow field structure and energy transfer efficiency. For monatomic gases (such as evaporated metal atoms), the hard-sphere model within the Chapman-Enskog theory approximates intermolecular collision behavior, as expressed by

$$\mu = \frac{\sqrt{mk_B T}}{d^2\pi^{3/2}}, \lambda = \frac{25}{32}\frac{\sqrt{\pi m k_B T}}{\pi d^2}\frac{k_B}{\left(\gamma-1\right)m} \tag{12}$$

For diatomic gases (such as $N_2$ and $O_2$ in air), the Lennard-Jones potential must be introduced to describe the long-range attractive and short-range repulsive interactions between molecules, thereby correcting the calculation of transport coefficients:

$$\mu = \frac{5}{16\sqrt{\pi}} \frac{\sqrt{mk_B T}}{\sigma^2 \Omega(T)}, \lambda = (\hat{c}_p + \frac{5}{4}\frac{R}{M})\mu \tag{13}$$

Here, $\sigma$ denotes the effective intermolecular collision diameter, and $\Omega(T)$ represents the collision integral.

For multicomponent mixtures, the dimensionless coupling factor $\phi_{\alpha\beta}$ corrects for the influence of ratios between molecular masses and viscosities of different components. The resulting equivalent viscosity coefficient and thermal conductivity can be expressed as

$$\mu_{\text{mix}} = \sum_{\alpha=1}^{N} \frac{x_\alpha \mu_\alpha}{\sum_\beta x_\beta \phi_{\alpha\beta}}, \quad \lambda_{\text{mix}} = \sum_{\alpha=1}^{N} \frac{x_\alpha \lambda_\alpha}{\sum_\beta x_\beta \phi_{\alpha\beta}}, \quad \phi_{\alpha\beta} = \frac{1}{\sqrt{8}}\left(1+\frac{M_\alpha}{M_\beta}\right)^{-\frac{1}{2}}\left[1+\left(\frac{\mu_\alpha}{\mu_\beta}\right)^{\frac{1}{2}}\left(\frac{M_\beta}{M_\alpha}\right)^{\frac{1}{4}}\right]^2 \tag{14}$$

Here, $x_\alpha$ and $x_\beta$ denote the mole fractions of the components, $\mu_\alpha$ and $\lambda_\alpha$ represent the viscosity and thermal conductivity of individual components, and $M_\alpha$ and $M_\beta$ correspond to the molar masses of the respective components.

## 2.5 Spectral Calculation Based on the Radiative Transfer Model

After solving for plasma expansion and ionization distribution, the relatively low plasma temperature ($10^3$~$10^5$ K) renders radiative energy loss negligible compared to particle kinetic and thermal energies. Consequently, radiation treatment serves as a post-processing step in flow field calculations. In this process, we obtain the spectral intensity $I$ distribution by solving the radiative transfer equation:

$$\frac{\partial I(z;v)}{\partial \tau} = -I(z;v) + S(z;v), \tag{15}$$

Here, $\tau$ denotes the optical thickness, and $S(z, v)$ represents the source function (under LTE conditions, the source function corresponds to the Planck blackbody function).

Integrating Eq. (15) yields a recursive formulation for the spatial transport of spectral intensity:

$$I_{v_i} = I_{v_{i-1}} e^{-\kappa \mathrm{d}z} + I_{v_i}^B \left(1 - e^{-\kappa \mathrm{d}z}\right), \quad i = 1, 2, ..., N \tag{16}$$

Here, $I_{vi}$ denotes the radiation intensity at the boundary of the $i$-th grid cell; $\kappa$ represents the total absorption coefficient corresponding to this frequency; d$z$ indicates the spatial grid size; and $I_{vi}^B$ is the blackbody radiation source term determined by the temperature of the grid cell.

According to Kirchhoff's law, the line spectral absorption coefficient $\kappa_{bb}$ is defined as

$$\kappa_{bb}(\nu)=\frac{\pi e^2 f}{m_e c} g_{low} n_i \frac{e^{-E_{low}/k_B T}}{Z_i(T)}\left(1-e^{-h\nu/k_B T}\right)P(\nu;\boldsymbol{\Gamma};\sigma) \tag{17}$$

Here, $f$ denotes the oscillator strength; $g_{low}$ represents the degeneracy of the lower energy level; $n_i$ is the ion number density; $E_{low}$ indicates the energy of the lower level; $Z_i(T)$ stands for the partition function; and $P(\nu)$ is the Voigt line profile function of the spectral line. This function is defined as the convolution of the Lorentzian broadening function at the line center ($L(\nu;\Gamma)$, due to Stark broadening) and the Gaussian broadening function in the far wings of the line ($G(\nu;\sigma)$, due to instrumental broadening):

$$P(\nu;\Gamma;\sigma)=\int_{-\infty}^{\infty} d\nu' L(\nu;\Gamma)G(\nu-\nu';\sigma) \tag{18}$$

Continuum radiation arises from two contributions: free-free bremsstrahlung (dominant in the long-wavelength infrared region) and free-bound radiative recombination (dominant in the short-wavelength ultraviolet region). These processes describe, respectively, the interaction between electrons and the Coulomb field of background ions, and the capture of free electrons by ions to form bound states with concomitant photon emission. Based on Kramers' theory, their expressions are given as follows[43]

$$\kappa_{ff}(\nu)=\frac{32\pi^3}{3\sqrt{3}}\frac{e^6(\hbar c)^2}{m_e c^2}\frac{1}{(2\pi m_e c^2 k_B T)^{1/2}}\bar{Z}^2 n_e n_i \frac{1}{(\hbar\nu)^3} g_{ib}(\nu)\left(e^{h\nu/k_B T}-1\right) \tag{19}$$

Here, $\bar{Z}$ denotes the average charge number; $g_{ib}(\nu)$ represents the Gaunt factor, which corrects for quantum effects in the classical derivation:

$$\begin{aligned}\kappa_{fb}(\nu)=&\frac{32\pi^3}{3\sqrt{3}}\frac{e^6(\hbar c)^2}{m_e c^2}\frac{1}{(2\pi m_e c^2 k_B T)^{1/2}}\bar{Z}^2 n_e n_i \frac{1}{(\hbar\nu)^3}\xi_z(\nu)\\ &\times e^{h\nu/k_B T}\left(1-e^{-h\nu/k_B T}\right)^2\end{aligned} \tag{20}$$

Here, $\xi_z(\nu)$ is a factor correcting for atomic electronic structure effects, taken as 1 within the temperature and photon frequency domains considered in this study; other symbols retain the meanings defined in Eq. (19).

## 2.6 Computational Domain and Numerical Framework

This study selects pure iron as the computational object. Table 1 lists the relevant physical properties, which provide the necessary basis for calculating ablation evaporation and ionizing radiation. The ionization calculations account for the first three ionization stages of pure iron, excluding background gas breakdown. In most laser processing, measurement, or diagnostic applications, background gas ionization is undesirable. Such ionization causes part of the laser energy to be absorbed by the background gas and converted into thermal energy. This reduces the effective laser energy incident on the target, resulting in energy loss.

**Table 1** Physical properties of pure iron used for calculations.

| Physical Property Parameter | Symbol/Unit | Value |
|---|---|---|
| Solid (Liquid) Density | $\rho_i$/(kg·m$^{-3}$) | 7874(6980) |
| Solid (Liquid) Specific Heat Capacity | $c_i$/(J·kg$^{-1}$·K$^{-1}$) | 449(507) |
| Solid (Liquid) Thermal Conductivity | $\lambda_i$/(W·m$^{-1}$·K$^{-1}$) | 80.2(34.5) |
| Enthalpy of Fusion | $\Delta H_{\mathrm{m}}$/(kJ·mol$^{-1}$) | 14 |
| Enthalpy of Vaporization | $\Delta H_{\mathrm{e}}$/(kJ·mol$^{-1}$) | 350 |
| Melting Point | $T_{\mathrm{m}}$/K | 1808 |
| Boiling Point | $T_{\mathrm{b}}$/K | 3023 |
| Solid-Liquid Coexistence Interval Width | Δ/K | 30 |
| Reflectivity (@1064 nm) | $R$ | 0.6602 |
| Optical Absorption Coefficient (@1064 nm) | $\alpha$/(10$^7$ m$^{-1}$) | 5.1043 |
| Atomic Diameter | $d_{\mathrm{Fe}}$/pm | 252 |
| First/Second/Third Ionization Energy | $E_i$/(kJ·mol$^{-1}$) | 762.5/1562/2957 |

For laser parameter settings, this study selects a Gaussian pulsed laser with a peak irradiance of 0.2 GW/cm$^2$ and a pulse width of 10 ns, assuming incidence at $t$ = 0. Regarding radiation calculations, the model covers the mid-ultraviolet and visible line and continuous spectra within the 180~830 nm range. Specifically, the line spectrum calculation includes the top 170 Fe I lines and 96 Fe II lines from the NIST Atomic Spectra Database. These lines fall within the 180~830 nm range, possess accuracy grades of B or higher, and exhibit high relative intensities. The specific list of spectral lines and their atomic transition properties is detailed in the Supplementary Materials.

In designing the numerical calculation framework, distinct spatiotemporal propagation schemes are applied to the target material and atmosphere regions. The target material region involves heat conduction, mass diffusion, and phase changes following laser-solid interaction. Consequently, this region requires solving unsteady convection-diffusion equations. Given the relatively smooth mass transport and the high precision requirements, a three-level fully implicit compact difference scheme[44] is selected. This scheme offers second-order accuracy in time and fourth-order accuracy in space. Furthermore, its unconditional stability allows for accurate capture of detailed variations during heat transfer and mass diffusion, while avoiding numerical instability.

Unlike the target material region, the atmosphere region involves the coupled effects of strong nonlinear transport, viscous diffusion, and radiation loss. It also includes calculations for shock waves and large flow gradients, necessitating a more robust and conservative numerical scheme. This study employs the Mac-Cormack explicit predictor-corrector scheme. Within each time step, the average time derivative of the flux variable $U$ to be solved equals the arithmetic mean of the predicted and corrected values:

$$\left(\frac{\partial U}{\partial t}\right)_{\text{av}} = \frac{1}{2}\left[\left(\frac{\partial U}{\partial t}\right)_i^t + \left(\frac{\partial \bar{U}}{\partial t}\right)_i^{t+\Delta t}\right] \tag{21}$$

This method exhibits low numerical dissipation. It is suitable for solving unsteady Navier-Stokes equations that contain both parabolic and elliptic terms. Furthermore, it achieves second-order accuracy in both time and space.

To further ensure computational stability, an additional fourth-order artificial viscosity correction term $\bar{S}_i^{t+\Delta t}$ is introduced. This term automatically smooths regions with shock waves and excessive flow gradients:

$$U_i^{t+\Delta t} = U_i^t + \left(\frac{\partial U}{\partial t}\right)_{\text{av}} \Delta t + \bar{S}_i^{t+\Delta t} \tag{22}$$

$$\bar{S}_i^{t+\Delta t} = \frac{C_x \left|\bar{P}_{i+1}^{t+\Delta t} - 2\bar{P}_i^{t+\Delta t} + \bar{P}_{i-1}^{t+\Delta t}\right|}{\bar{P}_{i+1}^{t+\Delta t} + 2\bar{P}_i^{t+\Delta t} + \bar{P}_{i-1}^{t+\Delta t}}\left(\bar{U}_{i+1}^{t+\Delta t} - 2\bar{U}_i^{t+\Delta t} + \bar{U}_{i-1}^{t+\Delta t}\right) \tag{23}$$

Here, $C_x$ is a dimensionless coefficient that regulates the intensity of artificial viscosity. Its value is determined based on the principles of balancing shock structure resolution and computational stability. Through parameter sensitivity analysis, this study selects $C_x = 0.2$.

This study also employs distinct meshing strategies for the computational domain and spatiotemporal settings. For spatial discretization, the target material region has a total length of 37.5 μm with a uniform grid size of 25 nm. The ambient gas region spans 1.5 cm with a grid size of 10 μm. Regarding temporal settings, the total simulation duration is 1 μs with a time step of 0.1 ns. This configuration satisfies the CFL stability criterion for the explicit time-marching scheme while ensuring accurate capture of the rapid thermal and flow effects during laser ablation.

# 3 Results and Discussion

## 3.1 Laser Energy Deposition and Ablation Process

Figure 1(a) illustrates the temporal evolution of laser irradiance at the material surface under laser pulse irradiation. The black curve represents the initial incident laser irradiance $I_{\text{input}}$, while the blue curve denotes the laser irradiance $I_{\text{target}}$ actually reaching the material surface. As shown in Figure 1(a), $I_{\text{input}}$ and $I_{\text{target}}$ nearly coincide during the initial stage. This indicates that plasma shielding is insignificant in the early phase of laser incidence because the plasma is still in its growth stage. Consequently, nearly all laser energy effectively reaches the material surface. However, $I_{\text{target}}$ decreases markedly as time progresses. Specifically, after ~15 ns, $I_{\text{target}}$ falls significantly below $I_{\text{input}}$. This reduction occurs primarily because the gradually forming plasma layer enhances the shielding effect on the incident laser, thereby reducing the laser energy received by the material surface. Figure 1(b) presents the spatiotemporal evolution of temperature within the material in an air atmosphere. The temperature near the material surface rises sharply after laser irradiation and gradually propagates inward. A distinct solid-liquid interface (*l-s* interface) emerges at approximately 8 ns. Just before the plasma shielding effect becomes significant at ~15 ns, the maximum internal temperature approaches 6700 K. As time advances, evaporation effects gradually dominate, and the gas-liquid interface (*g-l* interface) moves progressively into the material interior.

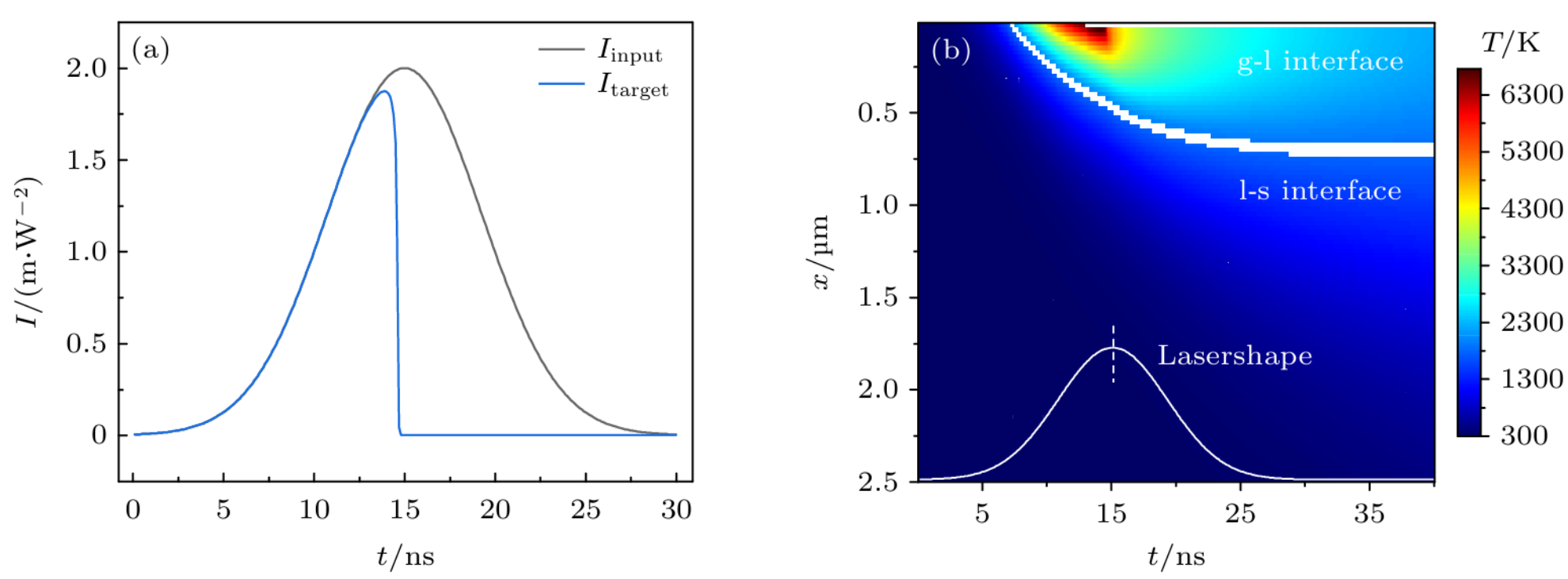

**Fig. 1** (a) Temporal evolution of laser irradiance on the material surface; (b) spatiotemporal evolution of temperature within the material.

## 3.2 Transport Processes at the Gas-Liquid Interface

Having clarified energy deposition and temperature response during the ablation phase, this section focuses on the gas-liquid interface as a transition region. We further analyze the transient evolution of transport characteristics at this interface by investigating the kinetic parameters of evaporating particles within the transition zone.

Figure 2(a) first illustrates the evolution of the liquid-phase ablation rate, $u_s$, during the initial 40 ns. During the onset of the laser pulse (approximately 0-13 ns), $u_s$ rises rapidly as incident laser energy accumulates. This corresponds to increasing surface temperature and intensified evaporation. However, a distinct gap appears in the $u_s$ curve at approximately 15 ns. This timing precisely coincides with the moment when plasma shielding effects become significant, as characterized in Figure 1(a). Plasma absorption of laser energy causes a sharp drop in the actual energy reaching the target surface, $I_{target}$. This reduction in energy supply for sustaining evaporation leads to a sudden decrease in $u_s$. Subsequently, as the laser pulse energy decays and the plasma continues to expand, $u_s$ gradually declines. It reaches zero at approximately 18 ns, by which time $I_{target}$ has dropped to zero and the evaporation process has nearly ceased. The evolution of the surface saturated vapor pressure, $P_s$, is shown in Figure 2(b). Its trend aligns closely with that of $u_s$, exhibiting similar gap and zeroing phenomena at approximately 15 ns and 18 ns, respectively. This consistency validates the strong coupling among surface temperature, evaporation rate, and vapor pressure established by the Hertz-Knudsen model and Knudsen layer boundary conditions.

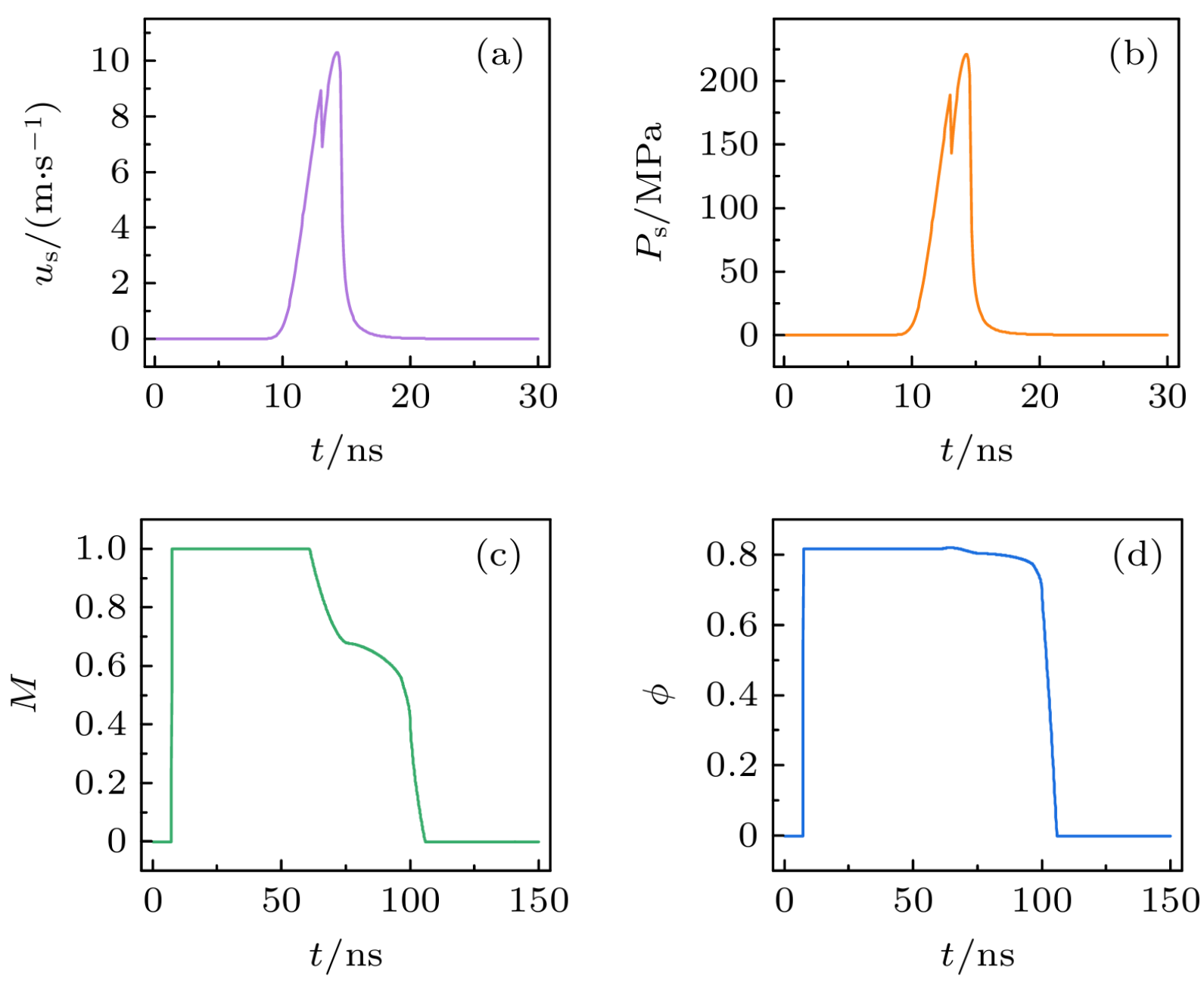

**Fig. 2** Evolution of interfacial transport characteristics under laser irradiation: (a) liquid-phase ablation velocity $u_s$; (b) surface saturated vapor pressure $P_s$; (c) Mach number $Ma$; (d) vapor mass flux across the Knudsen layer $\Phi$.

The kinetic state of the vapor outside the Knudsen layer is governed by the Mach number $M$, as shown in Fig. 2(c). During the initial evaporation stage, $M$ rapidly reaches 1, indicating that the vapor escapes from the gas-liquid interface at sonic velocity. This sonic flow regime persists for the majority of the effective evaporation duration. It only transitions to subsonic flow after approximately 60 ns, with $M$ dropping below 0.7 after ~70 ns. As a key free variable at the Knudsen layer boundary (Eq. (6)), $M$ cannot exceed 1. This numerical constraint arises because kinetic transitions across the Knudsen layer can only accelerate the vapor flow to the sonic state. In contrast, the vapor mass flow rate $\Phi$ in Fig. 2(d) quantitatively describes the intensity of mass transport across the layer. Throughout the period of Knudsen layer choking ($M = 1$), the mass flow of vapor escaping from the outer side accounts for 81.6% of the total evaporated mass. This implies that the vast majority of ablation products are transported to the vapor region under sonic expansion. Only 18.4% of the vapor remains within the non-equilibrium region of the layer. These results indicate that high-speed jetting, dominated by kinetic non-equilibrium effects, is the primary mode of mass transport during the early stage of nanosecond laser ablation.

## 3.3 Comparison of Hydrodynamic Results

After clarifying the initial transport characteristics at the gas-liquid interface, the ablation products enter the background atmosphere as high-temperature, high-pressure vapor. The subsequent expansion dynamics, energy evolution, and ionization states of the target particles constitute the core analysis of this section.

We first present the spatiotemporal evolution of plasma plume parameters. Figures 3(a) and (b) depict the distributions of plasma velocity $v_g$ and temperature $T_g$, respectively, over a timescale of 1 μs. We observe that after the laser pulse ends, the plume continues to propagate forward, accompanied by shock waves generated in the ambient medium. The shock front propagates to approximately 2 mm by ~400 ns. In the early stage of plume evolution, a high-temperature plasma core region exists near the material surface, expanding outward at high velocity. The maximum $v_g$ reaches approximately 6 km/s, while the maximum $T_g$ exceeds 90,000 K. The formation of this high-temperature, high-pressure core results from the continuous absorption of laser energy by the plasma during the ablation phase via inverse bremsstrahlung and photoionization mechanisms. It is further driven by its own pressure gradient during the subsequent expansion process. Meanwhile, the pressure $P_g$ in Fig. 3(c) and the plasma energy $E_{tg}$ in Fig. 3(d) reveal the energy storage and dissipation processes of the plume. The initial $P_g$ reaches the order of $10^8$ Pa, serving as the direct driving force for plume expansion. The evolution of $E_{tg}$ indicates that the system energy decays rapidly due to external work after the laser pulse ends.

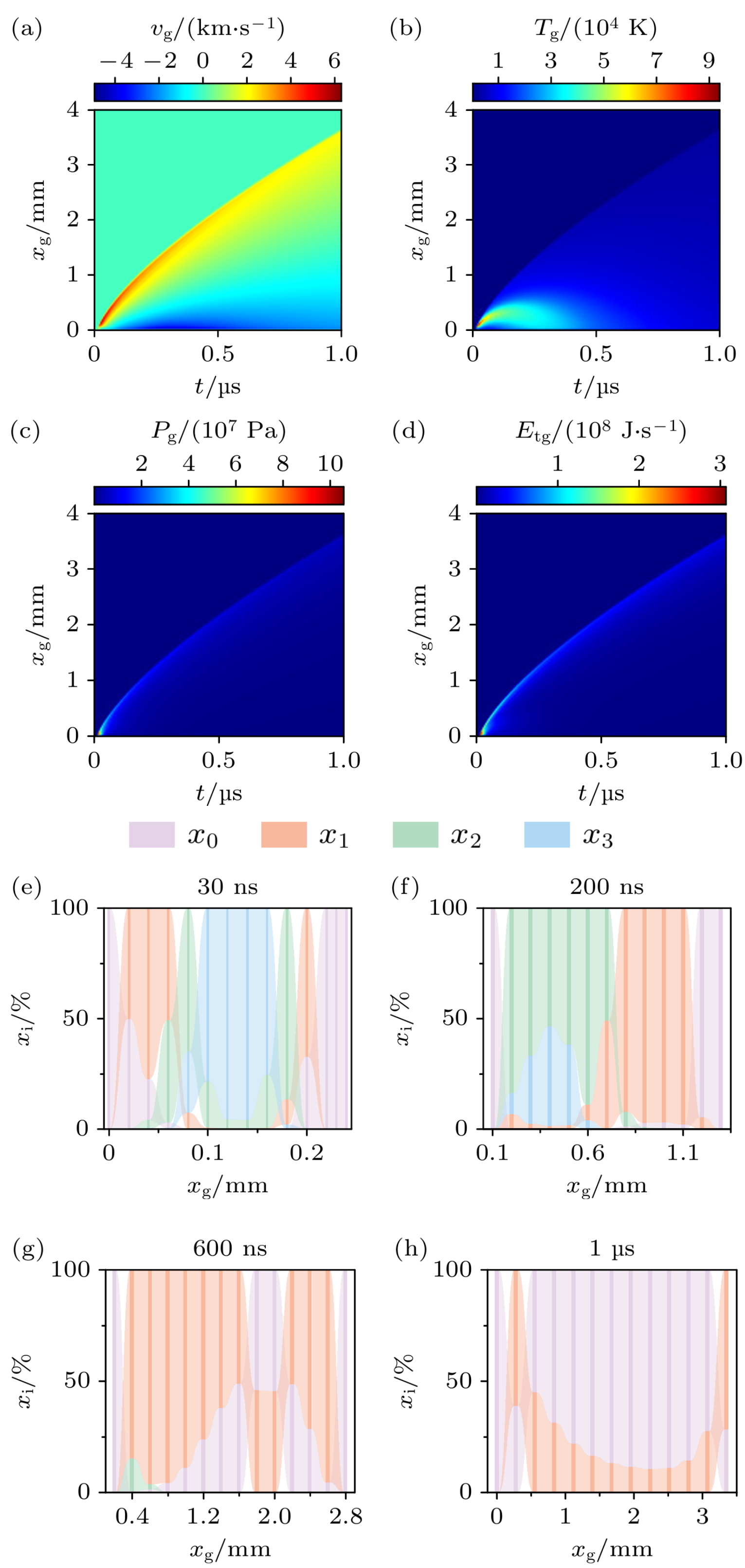

**Fig. 3** Spatiotemporal evolution of key parameters of the laser-induced plasma plume: (a) plasma velocity $v_g$; (b) plasma temperature $T_g$; (c) plasma pressure $P_g$; (d) plasma energy $E_{tg}$; spatial distributions of mass fractions in different ionization states for iron under an air atmosphere at (e) 30 ns, (f) 200 ns, (g) 600 ns, and (h) 1 μs.

The ionization state of the plasma is the decisive factor governing its radiative properties. Figures 3(e)-(h) illustrate the spatial distribution of the mass fractions of iron particles in various ionization states under an air atmosphere at different time points. During the early stage at 30 ns (Figure 3(e)), a spatial gradient structure emerges, transitioning from a high-temperature, high-ionization core to a low-temperature, low-ionization periphery. The plume core is dominated by high-valence $Fe^{3+}$ ions. This dominance corresponds to the extremely high local temperature (>50000 K) in this region, which causes the Saha equation (Eq. (10)) to shift strongly toward high-order ionization equilibrium. Subsequently, as the plume expands and cools, $Fe^{2+}$ becomes the dominant ion at 200 ns (Figure 3(f)). By 600 ns (Figure 3(g)) and 1 μs (Figure 3(h)), the plume core transitions from $Fe^{+}$ to being dominated by neutral iron atoms ($Fe^{0}$).

## 3.4 Comparison of Spectral Results

The radiative spectrum of the plasma is directly determined by its internal physical state and compositional information. Based on the radiative transfer model described in Section 2.5, this section further analyzes the temporal evolution of the spectrum during the expansion of laser-induced iron plasma. Figures 4(a)-(e) present the evolution of bremsstrahlung $I_{\mathrm{ff}}$, radiative recombination $I_{\mathrm{fb}}$, Fe I atomic line spectra $I_{\mathrm{FeI}}$, Fe II ionic line spectra $I_{\mathrm{FeII}}$, and the total radiation spectrum $I_{\mathrm{all}}$ (superposition of continuous and line spectra) at 400, 600, 800 ns, and 1 μs, respectively.

In the early evolutionary stage (400 ns), the plasma exists in a high-temperature, high-density state. Consequently, its radiation spectrum exhibits a strong continuous background dominated by ionic lines. Figures 4(a) and (b) illustrate the intensity distributions of $I_{\mathrm{ff}}$, which is dominated by long wavelengths, and $I_{\mathrm{fb}}$, which is dominated by short wavelengths, respectively. As shown in Figure 4(a), $I_{\mathrm{ff}}$ increases significantly with wavelength. It makes a prominent contribution in the long-wave band (500~700 nm). This trend aligns with Kramers theory (Eq. (19)). Complementarily, Figure 4(b) shows that $I_{\mathrm{fb}}$ increases sharply as the wavelength decreases. It dominates in the short-wave band (180~400 nm, the ultraviolet region). This occurs because $\kappa_{\mathrm{fb}}$ increases significantly when the photon energy exceeds the corresponding ionization energy threshold (Eq. (20)). As a result, high-energy photons released during the capture of free electrons by ions are concentrated mainly in the short-wave region. Superimposed on this continuous background, the line spectra comparisons in Figures 4(c) and (d) reveal that Fe II ionic lines are generally more intense than Fe I atomic lines. This observation is consistent with hydrodynamic calculations indicating a high degree of ionization in the plasma core at this stage (see Figure 4(e)).

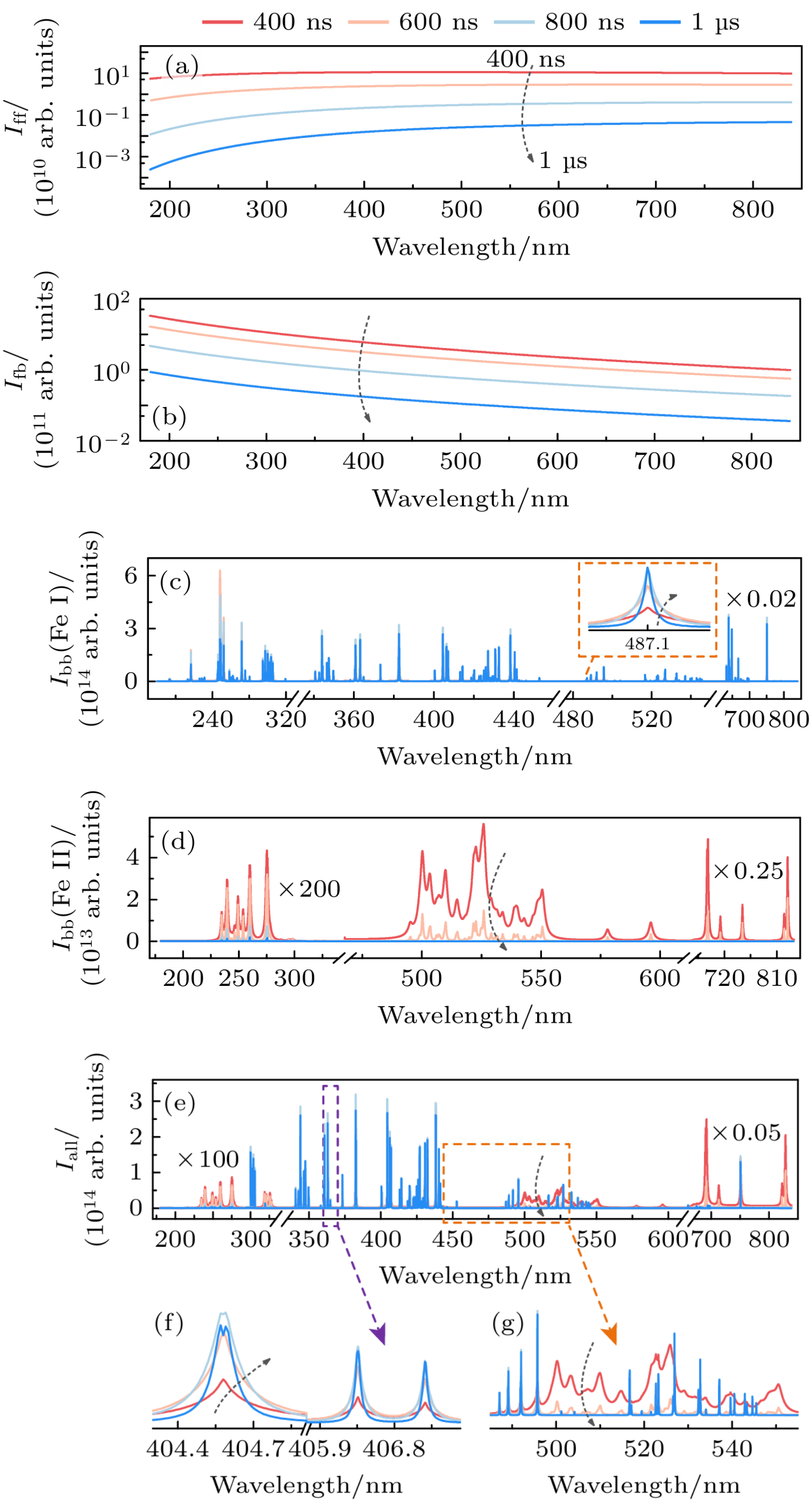

**Fig. 4** Evolution of the radiation spectra of laser-induced plasma: (a) Bremsstrahlung radiation $I_{ff}$; (b) radiative recombination $I_{fb}$; (c) Fe I atomic line spectrum $I_{FeI}$; (d) Fe II ionic line spectrum $I_{FeII}$; (e) total radiation spectrum $I_{all}$ combining continuum and line emissions; (f) local enlargement 1 of $I_{all}$; (g) local enlargement 2 of $I_{all}$.

As the plume expands and cools (600 ns-1 μs), the dominant spectral morphology gradually shifts. The intensity of continuous radiation decays significantly due to the decline in plasma temperature and total particle number density. Meanwhile, the relative intensity of Fe I atomic lines increases. This occurs because the decrease in plasma temperature shifts the Saha equilibrium toward neutral atoms (Fig. 3(e)-(h)). Consequently, the number density of $Fe^0$ increases while that of $Fe^+$ decreases, leading to enhanced atomic lines and weakened ionic lines.

Figures 4(f) and (g) highlight two characteristics of line spectrum evolution through local magnification of two spectral bands. In the range shown in Fig. 4(f), the temporal evolution of self-absorption in three typical Fe I lines (404.6, 406.4, and 407.2 nm) is observable. Between 400 and 600 ns, the overall intensity of these three lines increases, and their broadening expands. By 800 ns-1 μs, a dip appears in the peak region of the Fe I 404.6 nm line, exhibiting a self-reversal feature where the central intensity is lower than that of the wings. Simultaneously, the line width narrows. In contrast, the self-reversal features of Fe I 406.4 nm and Fe I 407.2 nm are less pronounced, showing only narrowed broadening and slightly lower intensity at 1 μs compared to 800 ns. This phenomenon occurs when the number density of neutral atoms in the cooler outer region of the plasma is sufficiently high to absorb characteristic photons emitted by the same element in the hot inner core. This self-absorption effect is a direct result of spatial non-uniformity in the plasma (presence of temperature and number density gradients). Its emergence and enhancement indicate that a sufficiently wide and cool outer region has formed in the plasma plume during the late stage of expansion.

Figure 4(g) illustrates the dynamic evolution of competitive dominance between Fe I and Fe II lines in the total radiation spectrum. Between 400 and 600 ns, this band is dominated by strongly broadened and overlapping Fe II lines, while Fe I lines are nearly submerged. This corresponds to the early state of the plasma, characterized by high temperature and high ionization degree. Over time, Fe II lines rapidly weaken and narrow due to decreased ionization and cooling. Subsequently, Fe I lines become prominent, emerging as relatively isolated and clear dominant lines in this band by 800 ns-1 μs. This reproduces the cooling process of the plasma plume transitioning from ion-dominated to atom-dominated.

### 3.5 Comparison with Experimental Results

After systematically analyzing the entire process of laser ablation, plasma evolution, and spectral radiation in an air atmosphere, this section compares the calculated spectra from this study with experimental measurements and results from other spectral calculation programs to verify the reliability of our calculations.

The experiment employed a nanosecond laser (Lapa 80, Raybow Optoelectronics; pulse energy ~12 mJ; spot diameter ~750 μm) focused directly to ablate a pure iron sample (YSBS11058-2021, NCS Testing Technology). Plasma spectra were collected using a ten-channel spectrometer (AvaSpec-ULS2048-10-USB2-RM Multi Channel, Avantes; spectral detection range 180-1060 nm; UF/UE gratings; average spectral resolution 0.05-0.18 nm). The detection delay was set to 400 ns, and the integration time was set to 1.05 ms (more specific experimental settings are available in our previous work[45]). To match the integration time of the experimental spectrometer, the transient spectra calculated in this study were integrated starting from 400 ns, with a step size of 100 ns, up to 4 μs (at which point the spectral intensity had decayed by seven orders of magnitude compared to 400 ns). Additionally, to establish a baseline for validation, the flow field calculation results were input into two spectral calculation programs for synchronous comparison: PrismSPECT (calculates plasma spectra including continuous emission and line transitions based on spatially averaged plasma parameters; one-dimensional planar configuration planar considering optical thickness and zero-thickness configuration zero-width assuming optical thinness were calculated separately; see Ref. [21] for detailed introduction) and NIST LIBS (calculates optically thin steady-state plasma line spectra under LTE at given temperatures and densities; see Ref. [25] for detailed introduction).

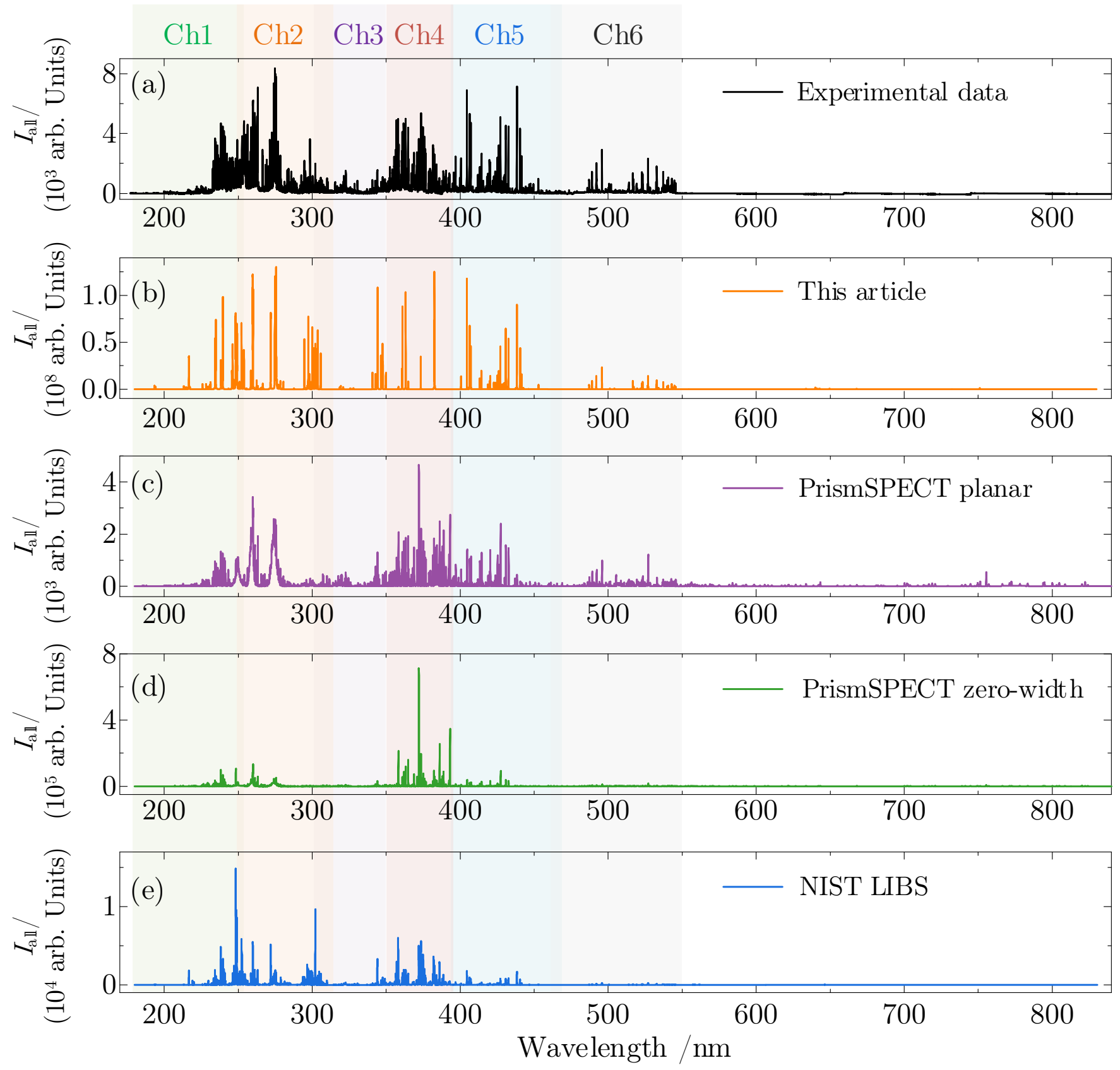

**Fig. 5** Comparison of time-integrated spectra for iron laser-induced plasma: (a) experimentally measured spectrum; (b) spectrum calculated using the model proposed in this study; (c) calculated using PrismSPECT planar; (d) calculated using PrismSPECT zero-width; (e) calculated using NIST LIBS. The shaded areas indicate the wavelength coverage of the spectrometer channels.

Figure 5 sequentially displays the experimentally measured spectrum (Fig. 5(a)), the spectrum calculated in this study (Fig. 5(b)), the PrismSPECT planar calculated spectrum (Fig. 5(c)), the PrismSPECT zero-width calculated spectrum (Fig. 5(d)), and the NIST LIBS calculated spectrum (Fig. 5(e)). The wavelength coverage ranges of the first six channels of the experimental spectrometer are also marked. Visually, the overall spectral morphology shows the highest similarity between the spectrum calculated in this study and the experimental spectrum. Both exhibit similar line distribution patterns across various bands. This is because the multi-physics coupling model established in this study self-consistently solves the radiative transfer equation, fully accounting for the spatial non-uniformity of plasma parameters and optical thickness. In contrast, although the PrismSPECT planar spectrum resembles the experimental spectrum in channels 1, 3, 5, and 6, the line broadening in channel 2 is excessive, and the relative intensities of lines in channel 4 deviate significantly from the experiment. This may stem from its partial consideration of radiative transfer under optical thickness, but its assumption of spatially averaged parameters fails to account for the actual temperature and density gradients between the plasma core and edges. Furthermore, PrismSPECT zero-width and NIST LIBS overestimate the intensity of resonance lines due to the neglect of optical thickness, resulting in the most significant differences in relative intensities across various bands compared to the experimental spectrum. The calculated spectrum in this study appears relatively sparse, primarily limited by the scale of atomic data used in the current model (170 Fe I lines and 96 Fe II lines). This can be improved in the future by expanding the number of calculated lines.

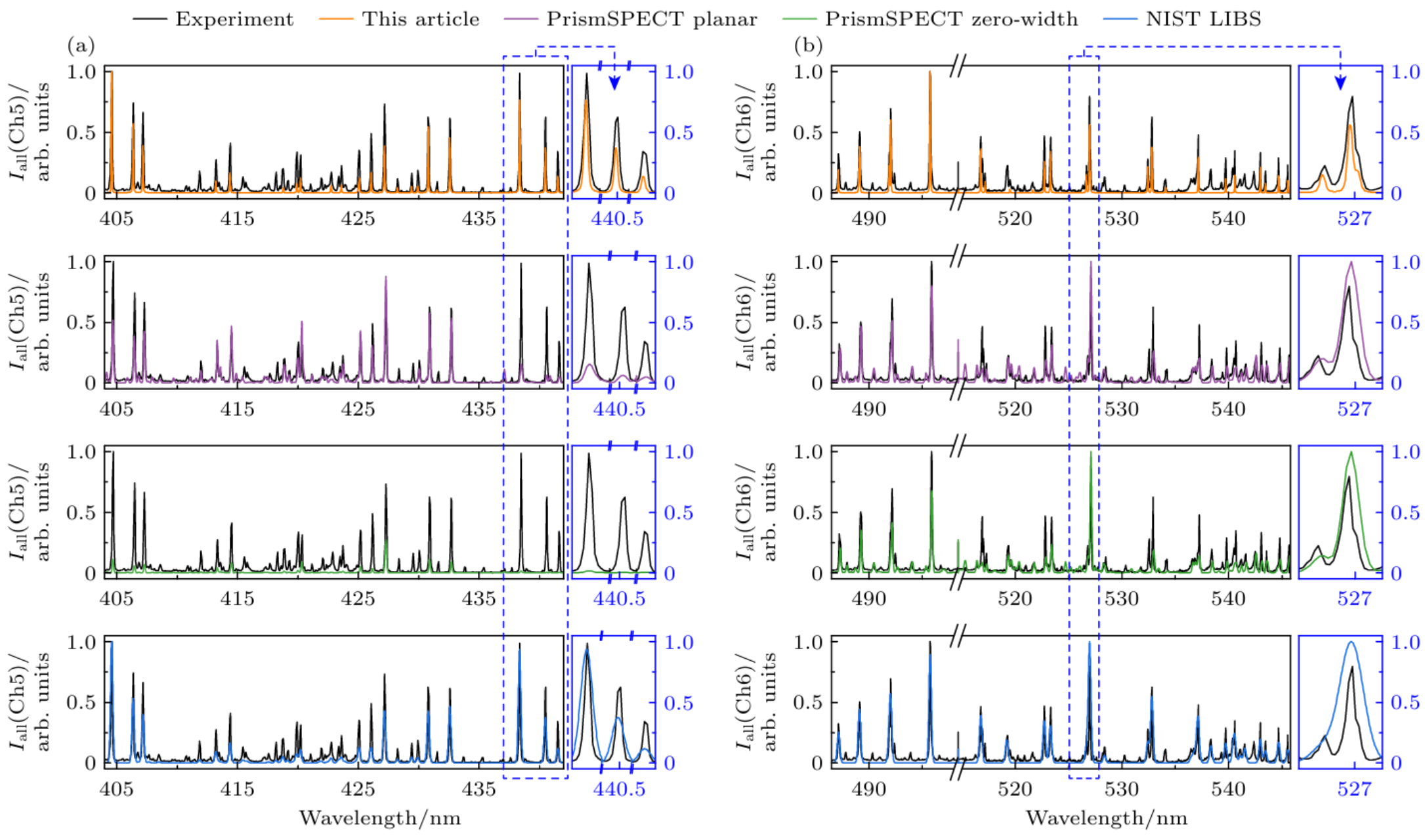

**Fig. 6** Local detailed comparison of two spectral channels: (a) Spectral comparison for Ch5 (390-480 nm); (b) spectral comparison for Ch6 (478-550 nm).

Figure 6 further presents a locally magnified comparison of spectrometer channel 5 (390-480 nm, Fig. 6(a)) and channel 6 (478-550 nm, Fig. 6(b)). It is evident that the spectrum calculated in this study not only matches the relative intensities of the experimental spectrum most closely but also demonstrates superior performance in wavelength resolution.

To quantitatively evaluate the performance of each computational method, this section introduces two statistical metrics for systematic assessment: $RSD_{a\text{-}a}$, based on the relative standard deviation of the ratio between experimental and calculated spectral line areas within each channel, and $RSD_{p\text{-}p}$, based on the relative standard deviation of peak ratios.

$$\mathrm{RSD}_{\text{a-a}}(\%)=\frac{\sqrt{\frac{1}{n-1}\sum_{i-1}^{n}\left(y_i^{\text{a-a}}-\bar{y}^{\text{a-a}}\right)^2}}{\bar{y}^{\text{a-a}}}\times 100\%,\quad y_i^{\text{a-a}}=\frac{\int_{\text{area}} I_i^{\text{simu}}\,\mathrm{d}\lambda}{\int_{\text{area}} I_i^{\text{exp}}\,\mathrm{d}\lambda} \tag{24}$$

$$\mathrm{RSD}_{\text{p-p}}(\%)=\frac{\sqrt{\frac{1}{n-1}\sum_{i-1}^{n}\left(y_i^{\text{p-p}}-\bar{y}^{\text{p-p}}\right)^2}}{\bar{y}^{\text{p-p}}}\times 100\%,\quad y_i^{\text{p-p}}=\frac{\max\left(I_i^{\text{simu}}\right)}{\max\left(I_i^{\text{exp}}\right)} \tag{25}$$

Here, $n$ denotes the total number of spectral lines in each channel, while  and  represent the experimental and simulated spectral intensities of the $i$-th line, respectively.

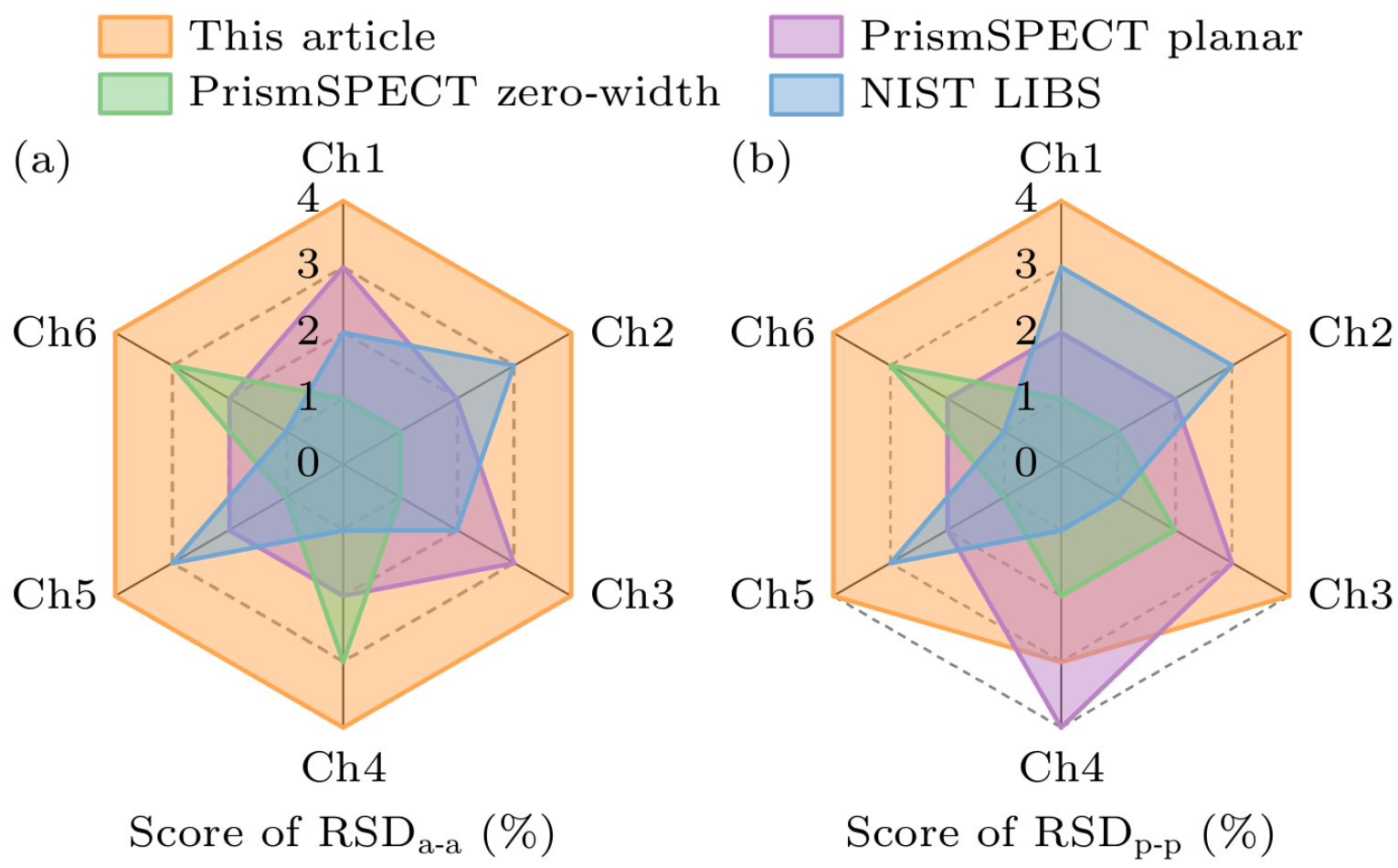

**Fig. 7** Quantitative assessment (radar chart) of the spectral prediction performance for different calculation methods: (a) Scores based on the relative standard deviation RSD of the spectral line area ratio ($RSD_{a\text{-}a}$); (b) scores based on the relative standard deviation of the spectral line peak ratio ($RSD_{p\text{-}p}$).

We then ranked and scored the $RSD_{a\text{-}a}$ and $RSD_{p\text{-}p}$ values for the four calculation methods in each channel from lowest to highest. A lower RSD indicates a better match between the simulated and experimental spectra, with scores ranging from 1 (lowest) to 4 (highest). The radar chart in Figure 7 shows that the proposed model achieved the highest comprehensive scores across nearly all channels and metrics. Specifically, for the $RSD_{a\text{-}a}$ metric (Figure 7(a)), our model received a perfect score of 4 in all six channels, significantly outperforming other methods (PrismSPECT planar scores: 3, 2, 3, 2, 2, 2; NIST LIBS scores: 2, 3, 2, 1, 3, 1; PrismSPECT zero-width scores: 1, 1, 1, 3, 1, 3). For the $RSD_{p\text{-}p}$ metric (Figure 7(b)), our model achieved the highest score in five of the six channels (scores: 4, 4, 4, 3, 4, 4), with only channel 4 scoring slightly lower than PrismSPECT planar.

# 4 Conclusion

This study develops a comprehensive, multi-physics coupled model spanning from laser energy deposition to spectral radiation. It systematically reveals the complete physical process of nanosecond laser interaction with an iron target. By self-consistently coupling phase-change ablation induced by laser energy deposition, the model accounts for the Knudsen layer interface under kinetic non-equilibrium, fluid ionization expansion, component diffusion, thermal/viscous transport, and spectral radiation transport. This approach achieves a unified description of complex physics across different regions and scales.

The research quantitatively characterizes the generation and evolution of the plasma shielding effect, linking it to the evaporation rate at the gas-liquid interface and the sharp drop in saturated vapor pressure. Furthermore, the Knudsen layer model clarifies that acoustic-speed expansion dominates the transport of early ablation products. It quantifies that the escaped vapor mass accounts for 81.6% of the total evaporated mass during this stage. As the formed plasma plume expands and gradually cools, its ionization state transitions from high-charge ions ($Fe^{3+}$) to neutral atoms ($Fe^{0}$). Based on this, spectral calculations demonstrate a dynamic transition in radiation characteristics. These shift from an early stage characterized by a strong continuous spectrum background and dominant ion lines to a later stage marked by continuous spectrum attenuation, prominent atomic lines, and the emergence of self-absorption. The appearance of self-absorption confirms that the model effectively captures the optical thickness effects arising from plasma spatial non-uniformity.

Through systematic comparisons with experimentally measured spectra and calculations from the PrismSPECT and NIST LIBS spectral codes, the proposed model achieves the highest comprehensive score in quantitative evaluations across multiple channels ($RSD_{a\text{-}a}$ and $RSD_{p\text{-}p}$). This validates the necessity and advantages of full-chain self-consistent modeling over traditional methods relying on spatial averaging or optically thin approximations for describing plasma non-uniformity and radiation transport. Consequently, it provides a numerical simulation framework for applications such as laser processing parameter optimization, quantitative analysis in laser-induced breakdown spectroscopy, and the design of novel plasma light sources.

## Acknowledgements

Project supported by the Joint Funds for Regional Innovation and Development of the National Natural Science Foundation of China (Grant No. U24A20154), the S&T Program of Energy Shaanxi Laboratory (Grant No. ESLB202414), and Key Research and Development Program of Shaanxi (Program No. 2024PT-ZCK-48).

## Author declarations

The authors have no conflicts of interest to disclose.

## Data availability

The data that support the findings of this study are available from the corresponding author upon reasonable request.